\documentclass{llncs}

\usepackage{color}
\usepackage{amsmath}
\usepackage{multirow}
\usepackage{booktabs}
\usepackage{url}
\usepackage{amsmath}
\usepackage{subfig}
\usepackage{adjustbox}
\usepackage{subfloat}
\usepackage{xcolor}
\usepackage{mdframed}
\usepackage{framed}
\usepackage{tabularx}
\usepackage{comment}

\usepackage{lipsum}
\usepackage{graphicx}

\usepackage{float}
\restylefloat{table}

\DeclareUnicodeCharacter{00B4}{'}

\newcommand\blfootnote[1]{%
  \begingroup
  \renewcommand\thefootnote{}\footnote{#1}%
  \addtocounter{footnote}{-1}%
  \endgroup
}

\begin{document}

\title{DemSelf, a Mobile App for Self-Administered Touch-Based Cognitive Screening: Participatory Design With Stakeholders}

\author{Martin Burghart\inst{1}\orcidID{0000-0002-8053-5253} \and Julie L. O´Sullivan\inst{2}\orcidID{0000-0002-8991-9966} \and Robert Spang\inst{1}\orcidID{0000-0001-6580-9060} \and Jan-Niklas Voigt-Antons\inst{1,3}\orcidID{0000-0002-2786-9262}}

\institute{Quality and Usability Lab, Technische Universit\"at Berlin, Berlin, Germany \\ \email{jan-niklas.voigt-antons@tu-berlin.de} \and Institut für Medizinische Soziologie und Rehabilitationswissenschaft, Charité - Universit\"atsmedizin Berlin, Berlin, Germany \and German Research Center for Artificial Intelligence (DFKI), Berlin, Germany}

\maketitle
\blfootnote{This paper has been accepted for publication in the Human-Computer Interaction International conference 2021. The final authenticated version is
available online at https://doi.org/[added after release of paper].}
\thispagestyle{empty}
\pagestyle{empty}

\begin{abstract}
Early detection of mild cognitive impairment and dementia is vital as many therapeutic interventions are particularly effective at an early stage. A self-administered touch-based cognitive screening instrument, called DemSelf, was developed by adapting an examiner-administered paper-based instrument, the Quick Mild Cognitive Impairment (Qmci) screen.\\
We conducted five semi-structured expert interviews including a think-aloud phase to evaluate usability problems. The extent to which the characteristics of the original subtests change by the adaption, as well as the conditions and appropriate context for practical application, were also in question. 
The participants had expertise in the domain of usability and human-machine interaction and/or in the domain of dementia and neuropsychological assessment. \\
Participants identified usability issues in all components of the DemSelf prototype. For example, confirmation of answers was not consistent across subtests. Answers were sometimes logged directly when a button is tapped and cannot be corrected. This can lead to frustration and bias in test results, especially for people with vision or motor impairments. The direct adoption of time limits from the original paper-based instrument or the simultaneous verbal and textual item presentation also caused usability problems. 
DemSelf is a different test than Qmci and needs to be re-validated. Visual recognition instead of a free verbal recall is one of the main differences. Reading skill level seems to be an important confounding variable.
Participants would generally prefer if the test is conducted in a medical office rather than at a patient's home so that someone is present for support and the result can be discussed directly.


\end{abstract}

\section{Introduction}
Currently there are no pharmacological treatments available for mild cognitive impairment (MCI) and most causes of dementia \cite{Petersen2014,Maier2016}. The diagnosis of potentially progressive cognitive impairment can cause fears and affected people often feel stigmatized and marginalized by society. These arguments can be put forward against an early diagnosis of MCI and dementia.
Nevertheless, in the view of most experts, the arguments for a diagnosis outweigh \cite{Knopman2014}: every patient has the right to be honestly informed about his or her health status. A diagnosis along with professional consultation can help family members better understand the behaviors they are experiencing. Early detection of MCI and dementia is vital as many therapeutic and preventive approaches – such as consultation, cognitive training, and physical exercise – are particularly effective at an early stage \cite{Petersen2014}. Such approaches can reduce the need for care and the burden on the patient and their caregivers.

Hence, there is a need for fast, reliable, and affordable screening instruments. The use of mobile devices such as tablets promises to better meet these requirements. Possible advantages include better standardization, automatic scoring, a digital test result, dynamic adaptation to the test person, or more accurate measurement of time and other factors. A person can also perform some tests without a trained medical professional, making cognitive testing less expensive and more accessible \cite{Bauer2012,Sternin2019}.
DemSelf is an attempt to adapt a validated paper-based instrument so that it can be performed independently on a touch-device. An expert evaluation provides insights into usability issues, differences to the original test and the appropriate context of use.

\section{Related Work}
For an average person, cognitive decline begins in the third decade and continues throughout life \cite{Mayr2012}. Cognitive aging affects domains such as attention, memory, executive functions, language, and visuospatial abilities \cite{Murman2015}.
Dementia is a brain syndrome associated with a decline of cognitive functioning that is "not entirely attributable to normal aging and significantly interferes with independence in the person’s performance of activities of daily living." \cite{HealthOrganization2020}. 
The most common form of dementia is due to Alzheimer's disease (AD), but dementia can occur in a number of different circumstances and diseases, such as vascular diseases, Parkinson's disease or Lewy body dementia \cite{Maier2016}.
Cognitive decline is the clinical hallmark in dementia and memory impairment is the most prominent symptom in most patients. A key feature of dementia is that everyday skills such as using public transport or handling money are affected. AD in particular has a progressive course. In severe dementia, patients are almost completely dependent on help from others.
In the course of research on cognitive aging, some people have been found to be in a "gray area between normal cognitive aging and dementia" \cite[p. 370]{Stokin2015}. 
A person shows cognitive decline that is beyond normal ageing, but functional activities of daily living are not affected. MCI is one of the most widely used and empirically studied terms to describe this state. A precise and universal definition of MCI has not yet been established \cite{Maier2016}; see \cite{Stokin2015} for a discussion on similar concepts such as Mild Neurocognitive Disorder.
In some cases, MCI can be a pre-clinical stage of dementia, particularly AD \cite{Maier2016}. Early identification of MCI would allow interventions at an early stage which may influence the course of the disease. However, people with MCI often remain undiagnosed \cite{Cordell2013}.\\
An early step in diagnosing cognitive impairment is often a brief cognitive screening instrument. 
The administration usually takes only a few minutes and allows assessment of different cognitive domains such as attention, memory, orientation, language, executive functions, or visuospatial abilities. 
A cognitive screening instrument provides information on the presence and severity of cognitive impairment in patients with suspected dementia or MCI. 
There is a multitude of different cognitive screening instruments. Recent systematic reviews and meta-analyses are available \cite{Tsoi2015,Aslam2018,Roeck2019}. The right choice depends on numerous factors: "Clinicians and researchers should abandon the idea that one screening instrument ... can be used in every setting, for all different neurodegenerative diseases and for each population." \cite[p. 11]{Roeck2019}.\newline Most available cognitive screening instruments are primarily paper-based and administered by healthcare professionals. However, computers, tablets, or similar devices can also be used to administer, score, or interpret a cognitive test. \\
Computerized cognitive screening instruments offer several potential advantages over paper-based instruments, such as increased standardization of scoring, ease of administering in different languages, reduced costs, remote testing, adaption, and more precise measurement of time- and location-sensitive tasks \cite{Bauer2012,Sternin2019}.
A self-administered web-based test Brain on Track, for example, uses random elements to minimize learning effects in longitudinal tests \cite{Ruano2016}.
Other computerized tests are designed as mini-games to achieve a more relaxed testing environment and to reduce the drop-out rates in longitudinal testing \cite{Zeng2018}, or use scoring algorithms similar to principal components analysis to improve sensitivity \cite{Shankle2005}.
Individual factors like technical experiences, attitudes towards technology, or non-cognitive impairments such as motor or sensory disabilities may affect interaction with the computer interface and bias the test result. A computerized screening instrument might therefore not be suitable for certain people.
Test developers should consider such factors during validation and report their influences on the test \cite{Bauer2012}.
Computerized cognitive assessment already has a long tradition and a variety of instruments is available \cite{Zygouris2014,Aslam2018,Roeck2019}. Nevertheless, they are still used much less frequently in practice than paper-based tests \cite{Sternin2019}. The lack of normative population data is one problem, making it difficult to choose the right instrument.\\

Usability is another crucial aspect in self-assessment by elderly users with possible cognitive and other impairments, posing a potential barrier to the practical use of self-assessment instruments.
Mobile devices with touch displays are often used for computer-based testing as they allow direct manipulation of information, which can feel very natural to the user \cite{DanielWigdor2011}. However, most applications do not take into account the needs of the elderly. Several age-related factors affect interaction with touch interfaces in mobile applications \cite{Balata2015,Fisk2004}. Older people can have trouble identifying thin lines, reading small and low-contrast text, or distinguishing between visually similar icons. Fine movements, time critical gestures (e.g. double taps), or complex multi-touch gestures can also be problematic.
Other studies have shown that mobile devices such as tablets can be successfully integrated into the lives of people with dementia - for example, to assess quality of life \cite{Junge2020}, to improve quality of live via cognitive training games or communication with staff and family members \cite{Cha2019,Antons2016}, or to improve outpatient dementia care by fostering guideline-based treatment \cite{Lech2019,Lech2018}.
Usability aspects should be considered early on in the development of a new instrument. Some authors have chosen cognitive training exercises as subtests where high usability has already been confirmed \cite{Ruano2016}. In general, however, the usability must be evaluated specifically for the new instrument. Various methods of expert-based evaluation and tests with users are available for this purpose. For example, some authors had organized focus groups with doctors and healthy older people and asked them to rate the usability of their instrument with a list of 12 statements based on the System Usability Scale \cite{Zeng2018}.
Usually, several cycles of re-design and evaluation are necessary to achieve high usability, since not all problems are discovered in one pass and a new design may also reveal or create new problems \cite{Moeller2017}.

\section{Methods}
We conducted five semi-structured interviews with usability and domain experts to determine which usability problems exist in the DemSelf app and how they could be solved. The extent to which the characteristics of the original subtests change by the adaption, as well as the conditions and appropriate context for practical application, were also in question.

\subsection{Participants}
The participants had expertise in the domain of usability and human-machine interaction and/or in the domain of dementia and neuropsychological assessment. The average age was 29.6 years (\textit{SD} = 1.34). All participants were female and had a university degree as their highest completed level of education. Three participants worked as research assistants in an MCI-research domain, while two participants were professionals in usability and user experience.
Participants had an average of 7.2 years (\textit{SD} = 2.5) of professional experience in human-machine interaction or usability. Three participants reported practical experience in the diagnosis of MCI and dementia. Only one participant reported practical experience with computerized digital screening instruments. All participants reported practical experience in assessing the usability of touch-based software. Two participants reported practical experience in assessing the usability of touch-based software specifically for elderly people.

\subsection{Procedure}
Four interviews were conducted in person. Participants operated the DemSelf app on an iPad (6th Generation). One interview was conducted remotely via telephone call. Here, the app was simulated in Xcode on an iPad (6th Generation) and operated remotely via TeamViewer. The average interview duration was 71 minutes.

Participants were informed about dementia, MCI, and the purpose of cognitive screening instruments. Each participant then completed the DemSelf testing process twice. We first presented a scenario in which an elderly patient is asked by a physician to perform the test alone. Participants were asked to put themselves in that person's position and to solve the test as correctly as possible. We encouraged participants to think aloud during the first round. This procedure is a form of cognitive walkthrough \cite{Mahatody2010} in the sense that a usage context, a user, and a task were specified. The goal was to uncover usability problems when the test is performed for the first time as this is the most common use of a screening instrument.\\
In the second round, participants could try alternative inputs and make typical mistakes. We asked participants to clarify comments made in the first round of thinking aloud and asked further questions regarding usability and differences between the original and the adapted subtests. After completing the second round, participants could comment on the adequate context of use and conditions for using the app as a cognitive screening instrument.

\subsection{Apparatus}
We developed a self-administered touch-based cognitive screening instrument, called DemSelf, by adapting an examiner-administered paper-based instrument, the Quick Mild Cognitive Impairment (Qmci) screen. The goal of DemSelf is to classify whether a person has normal cognition, MCI, or dementia. The test is to be performed independently on a mobile device such as a tablet.

The Qmci was selected based on the following criteria: high accuracy in detecting MCI, short administration time (under 10 minutes), and detailed instructions for administration and scoring. The Qmci was originally published in 2012 and is specifically designed to differentiate MCI from normal cognition \cite{OCaoimh2012a}. Two recent systematic reviews show reliable results for detecting MCI \cite{Glynn2019,Roeck2019}. It is a short (3–5 minute) instrument composed of six subtests – Orientation, Word Registration, Clock Drawing, Delayed Recall, Word Fluency, and Logical Memory. There are cut-off scores for MCI and dementia adjusted for age and education. The Qmci covers the cognitive domains orientation, working memory, visuospatial/construction, episodic memory and semantic memory/language \cite{OCaoimh2017}. 

We translated the items from the Qmci into German and based the instructions and scoring system of DemSelf on the Qmci guide \cite{Molloy2017}.  
In the Qmci there is a verbal interaction between the test administrator and the subject. In a self-assessment, speech recognition would come closest to this interaction, but was considered too error-prone. The Qmci subtest Word Fluency, in which as many animals as possible are to be be named in one minute, is therefore not included in DemSelf. Keyboard input was identified as a major difficulty for older people and a frequent source of errors in another self-assessment screening instrument \cite{Ruano2016}. Therefore, most user input in DemSelf is done by tapping large, labeled buttons that represent either correct answers or distractors. Thus, the answer choices are limited in DemSelf – this is one of the key differences from Qmci. Answer choices are randomly distributed for repeated testing. 

\subsubsection{Consent}
\label{subsubsec:consent}
\begin{figure}[ht]
\minipage[t]{0.3\textwidth}
  \includegraphics[width=\linewidth]{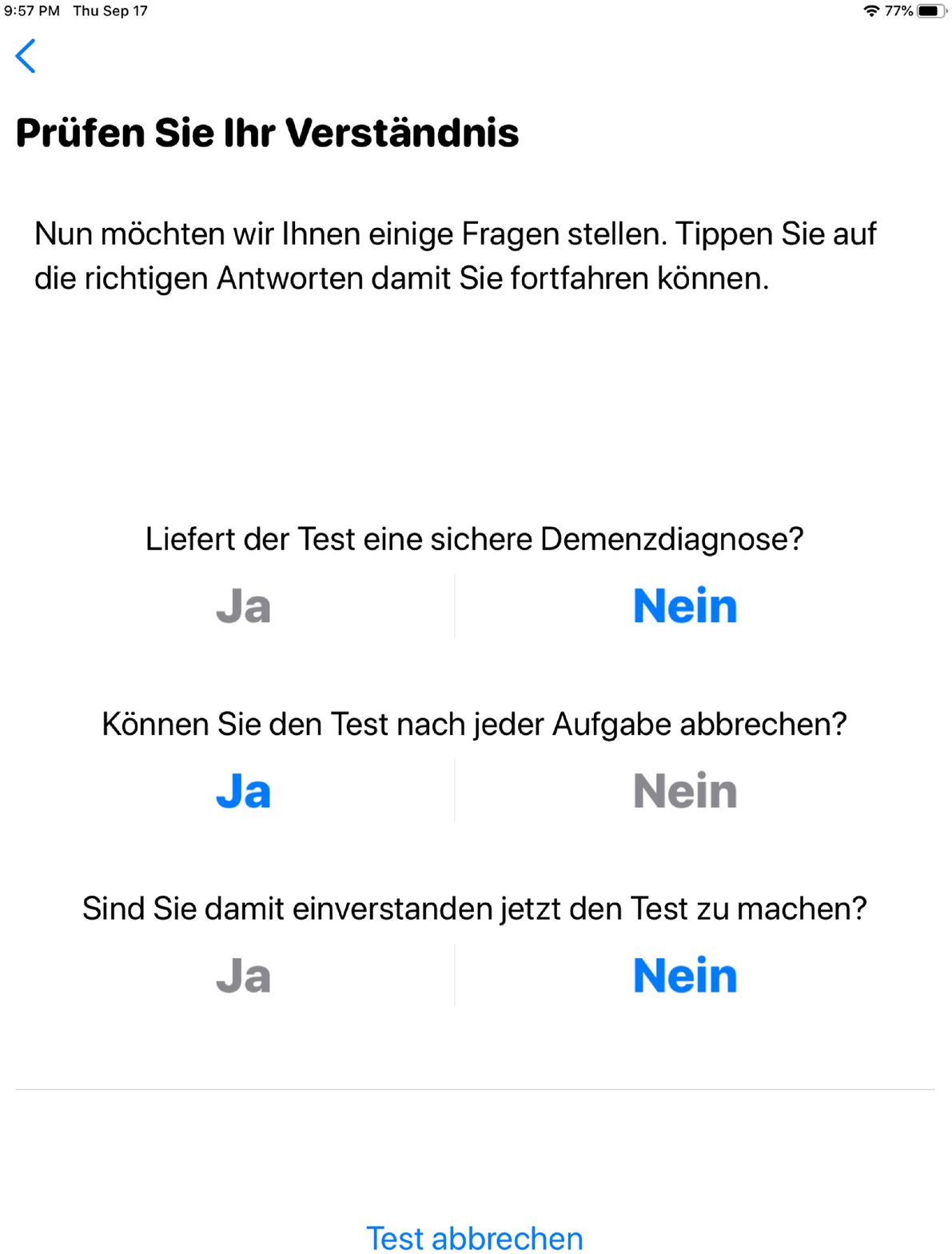}
  \begin{center}{\footnotesize(a) Check understanding of test limitations.\par}\end{center}
\endminipage\hfill
\minipage[t]{0.3\textwidth}
  \includegraphics[width=\linewidth]{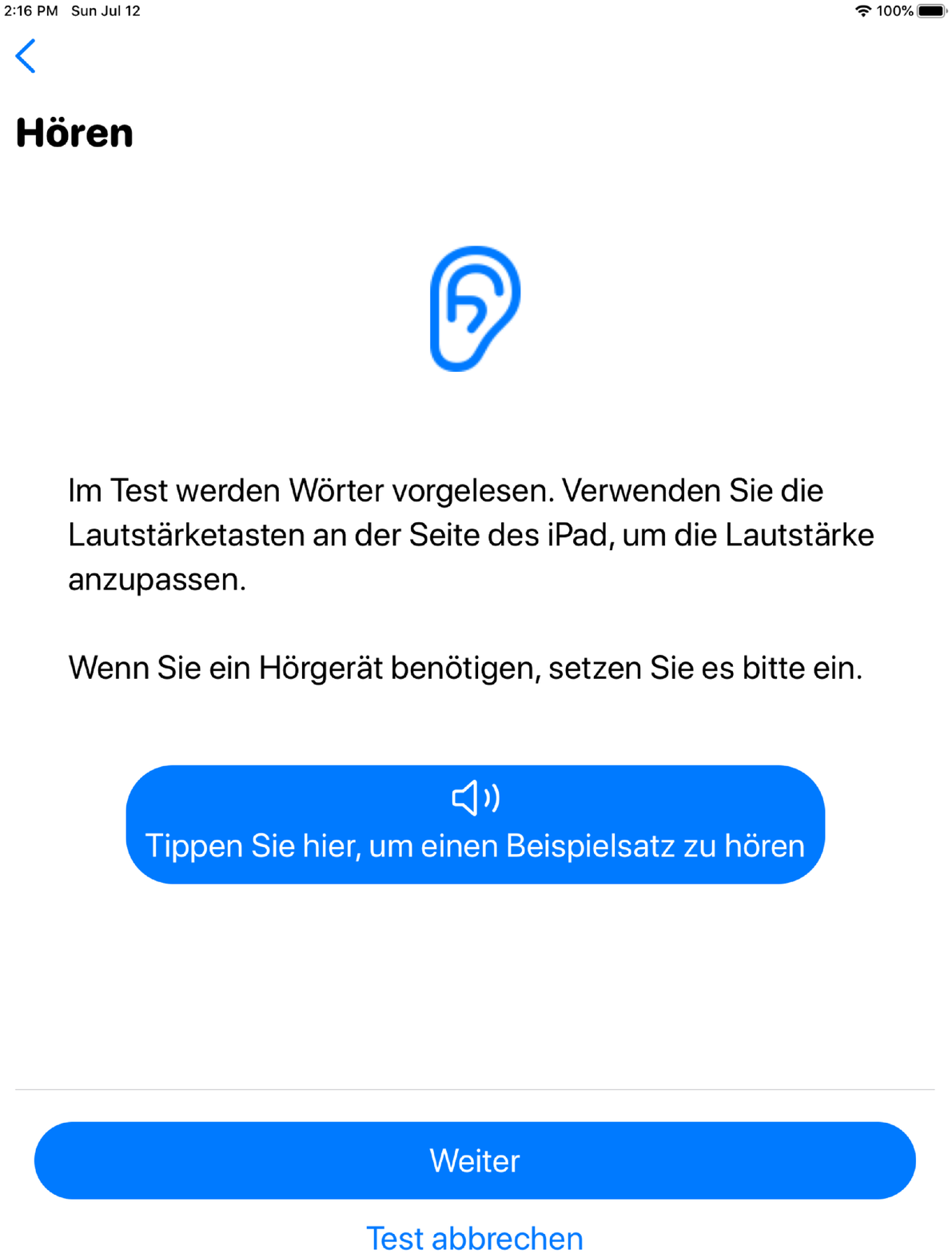}
  \begin{center}{\footnotesize(b) Test speech output.\par}\end{center}
\endminipage\hfill
\minipage[t]{0.3\textwidth}
  \includegraphics[width=\linewidth]{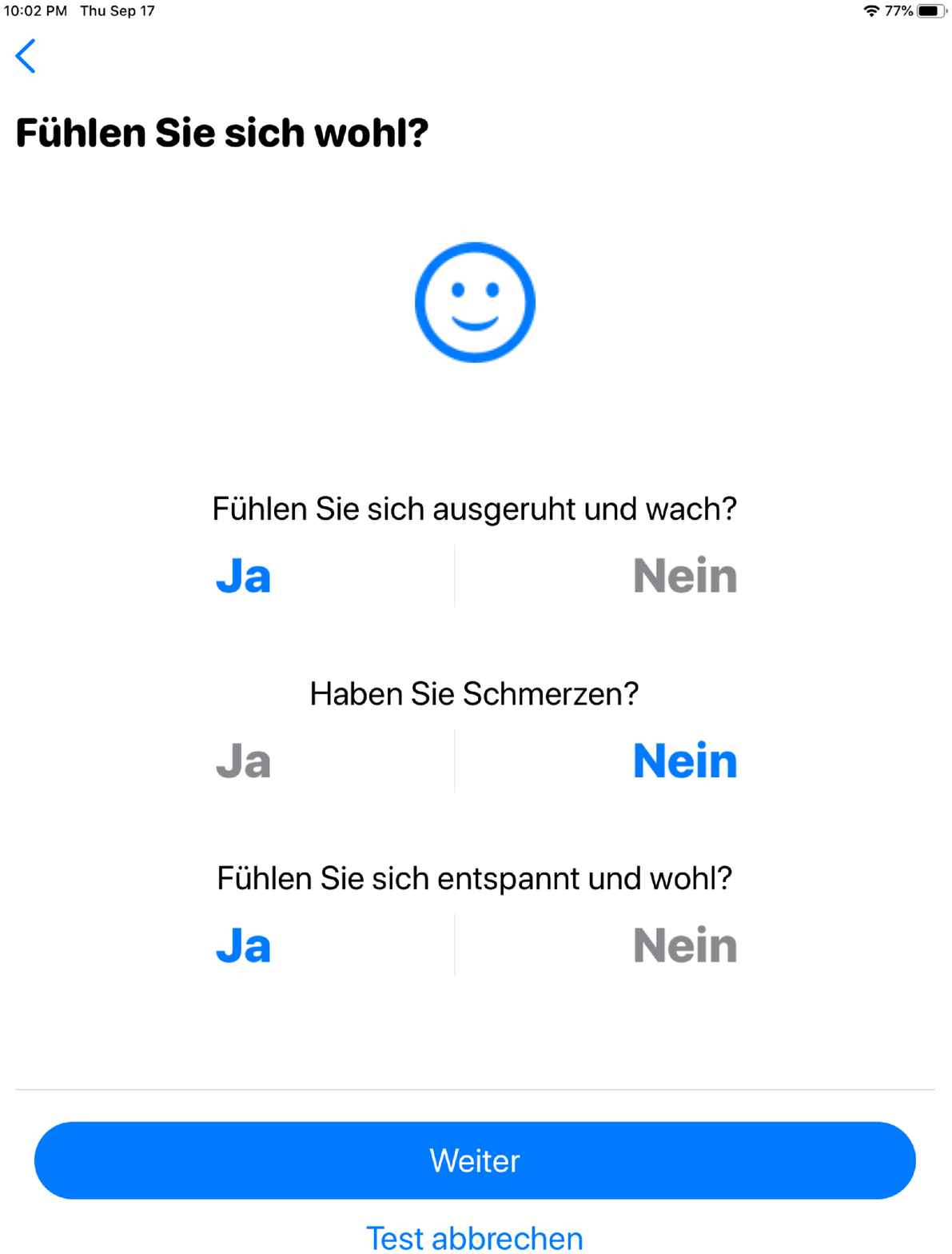}
  \begin{center}{\footnotesize(c) Check emotional and physiological state.\par}\end{center}
\endminipage
\caption{Selection of screens before the test begins. See Section \ref{subsubsec:consent} for a detailed description. Translation of instructions: (a) Now we would like to ask you some questions. Tap on the correct answers so that you can continue. Does the test provide a reliable diagnosis of dementia? Can you stop the test after each task? Do you agree to take the test now? (b) In the test, words are read aloud. Use the volume buttons on the side of the iPad to adjust the volume. If you need a hearing aid, please insert it. (c) Do you feel rested and awake? Are you in pain? Do you feel relaxed and comfortable?}
\label{fig:ConsentTestEnvironmentScreens}
\end{figure}

DemSelf first requires informed consent from the patient. To limit the demands on working memory and attention, each screen contains only a few short sentences with information about the risk of a false result, the data collected, the duration of the test, and the option to stop the test at any time. Subjects are encouraged to talk to a physician if they have any doubts about the test.
Any isolated cognitive screening instrument is insufficient for a diagnosis of MCI or dementia \cite{Aslam2018}. The Qmci guide recommends verbally clarifying the purpose and indicating that the subject can stop at any time \cite{Molloy2017}. In DemSelf, this comprehension is tested with the questions: Does the test provide a reliable diagnosis of dementia (No)? Can you stop the test after each task (Yes)? Subjects can continue only if both questions are answered correctly and if they agree to take the test (see Figure \ref{fig:ConsentTestEnvironmentScreens}).

\subsubsection{Test Environment}
The aim of cognitive testing is to achieve the best possible performance to ensure that deficits are not caused by internal or external confounding factors \cite{Pentzek2010}. The subject is therefore instructed to perform the test in a quiet environment without distractions and to use a hearing aid and wear glasses if necessary. The volume setting and the reading aloud function can be tested in advance. The physiological and emotional state of the person being tested can cause the test result to be biased \cite{Overton2016}. A subject is therefore asked whether he or she feels well, is not in pain, is not emotionally upset, and is not tired (see Figure \ref{fig:ConsentTestEnvironmentScreens}). The answers are reported in the test result for the healthcare professional.

\subsubsection{Subtest Orientation}
DemSelf asks five questions about the country, year, month, date (for the day), and day of the week. There are 12 answer choices for country and year including the correct answer and 11 distractors: 7 countries are randomly selected European countries (testing was assumed to take place in Germany). The remaining 4 countries are randomly selected from a list of all countries on earth. The distractors for the current year are randomly selected from a span of \(+/- 20\) years. For month, date, and day of the week all available options are given as possible answers. There is a 10 second time limit for answering before the next question appears.

\subsubsection{Subtest Word Registration}
\label{subsubsec:wordRegistration}
In Word Registration, DemSelf displays 5 words on the screen and consecutively reads each word aloud. The Swift class AVSpeechSynthesizer was used for speech synthesis with a German female voice. The speech rate was set to 0.40 throughout the app with a pause of \(1/2\) seconds before and after each utterance. The subject is then asked to tap the previously presented words in any order. The 16 answer options include 11 distractors which are randomly selected from a list of semantically and syntactically related words. If the subject does not select the correct words, the items are presented again up to 3 times and the subject must select the words again.

\begin{figure}[ht]
\minipage[t]{0.3\textwidth}
  \includegraphics[width=\linewidth]{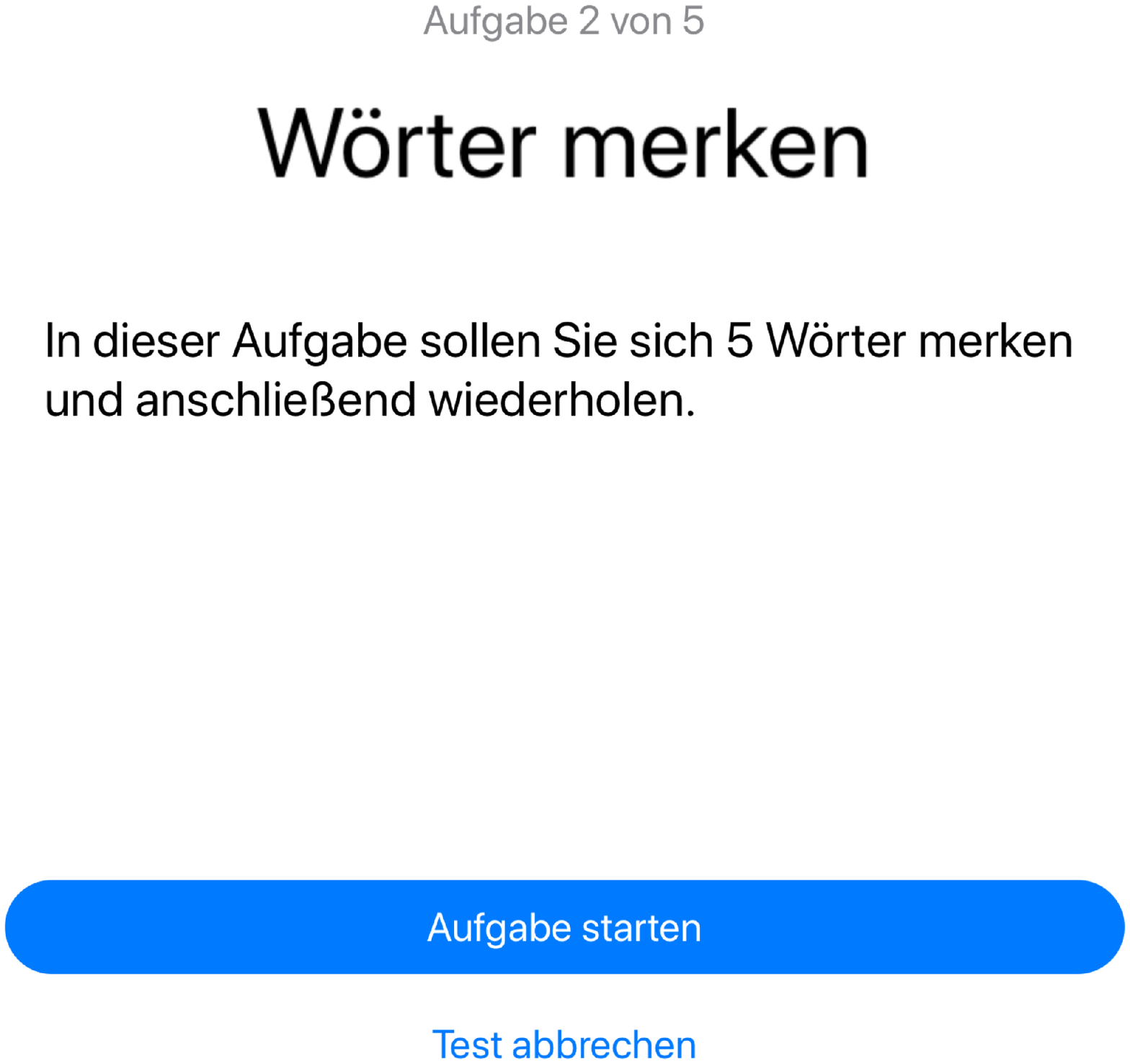}
  \begin{center}{\footnotesize(a) Introductory screen.\par}\end{center}
\endminipage\hfill
\minipage[t]{0.3\textwidth}
  \includegraphics[width=\linewidth]{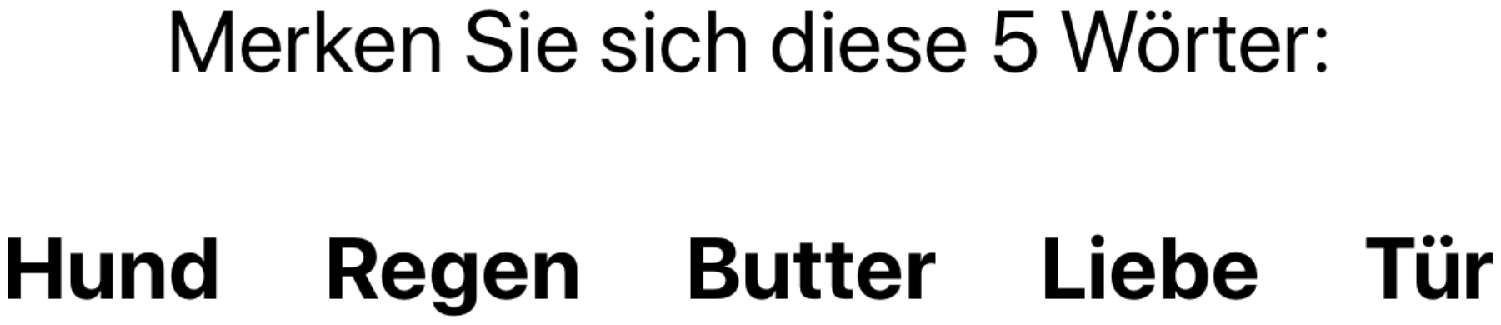}
  \begin{center}{\footnotesize(b) Items are displayed and read aloud.\par}\end{center}
\endminipage\hfill
\minipage[t]{0.3\textwidth}
  \includegraphics[width=\linewidth]{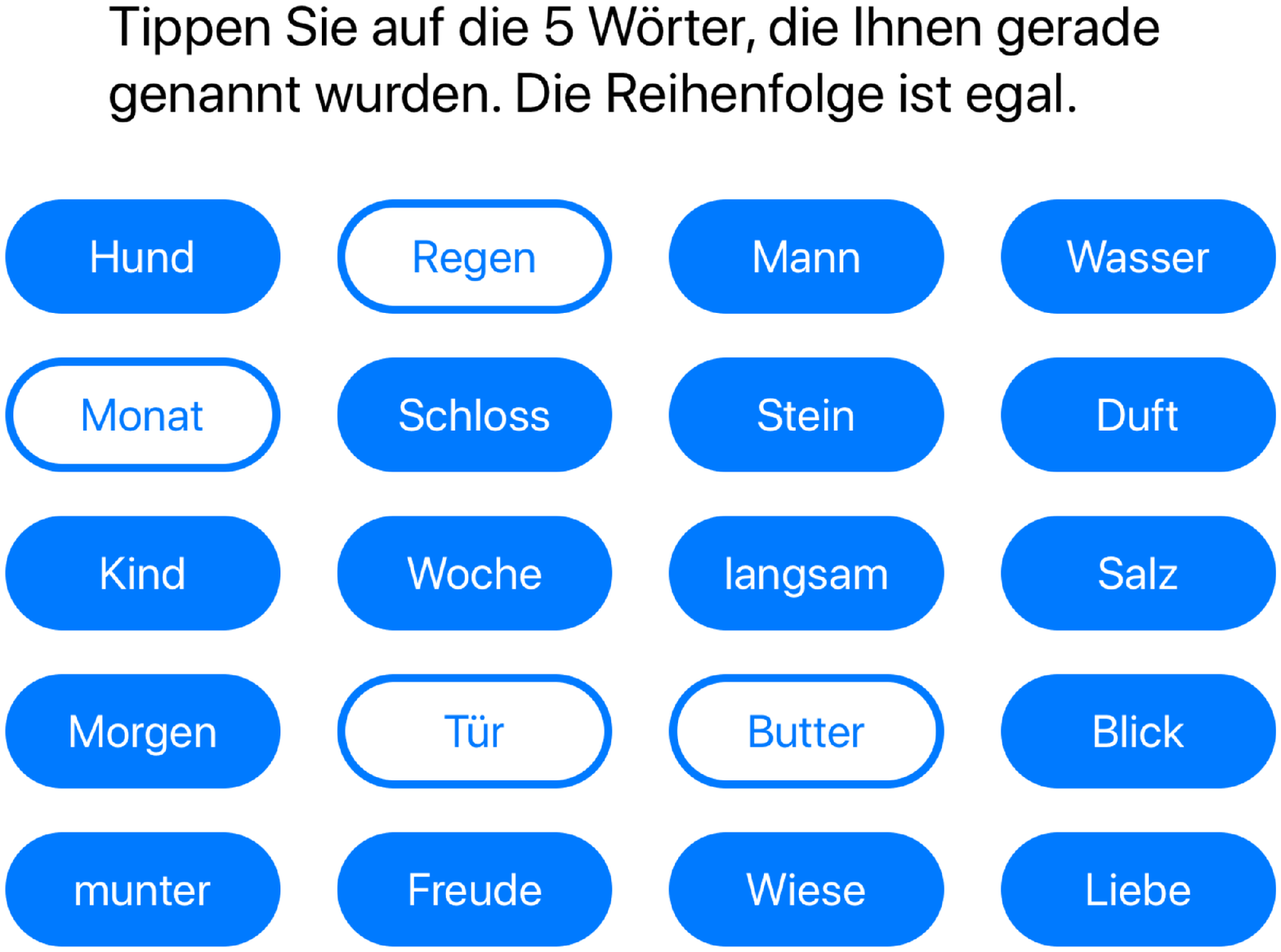}
  \begin{center}{\footnotesize(c) Answers are selected by tapping\par}\end{center}
\endminipage
\caption{Word Registration: item presentation and selection from different answer choices. See Section \ref{subsubsec:wordRegistration} for a detailed description. Translation of instructions: (a) In this task you are supposed to remember and then repeat 5 words. (b) Remember these 5 words. (c) Tap on the 5 words you were just told. The order does not matter.}
\label{fig:WordRegistration_input}
\end{figure}

\subsubsection{Subtest Clock Drawing}
\label{subsubsec:clockDrawing}
\begin{figure}[!htb]
\minipage[t]{0.3\textwidth}
  \includegraphics[width=\linewidth]{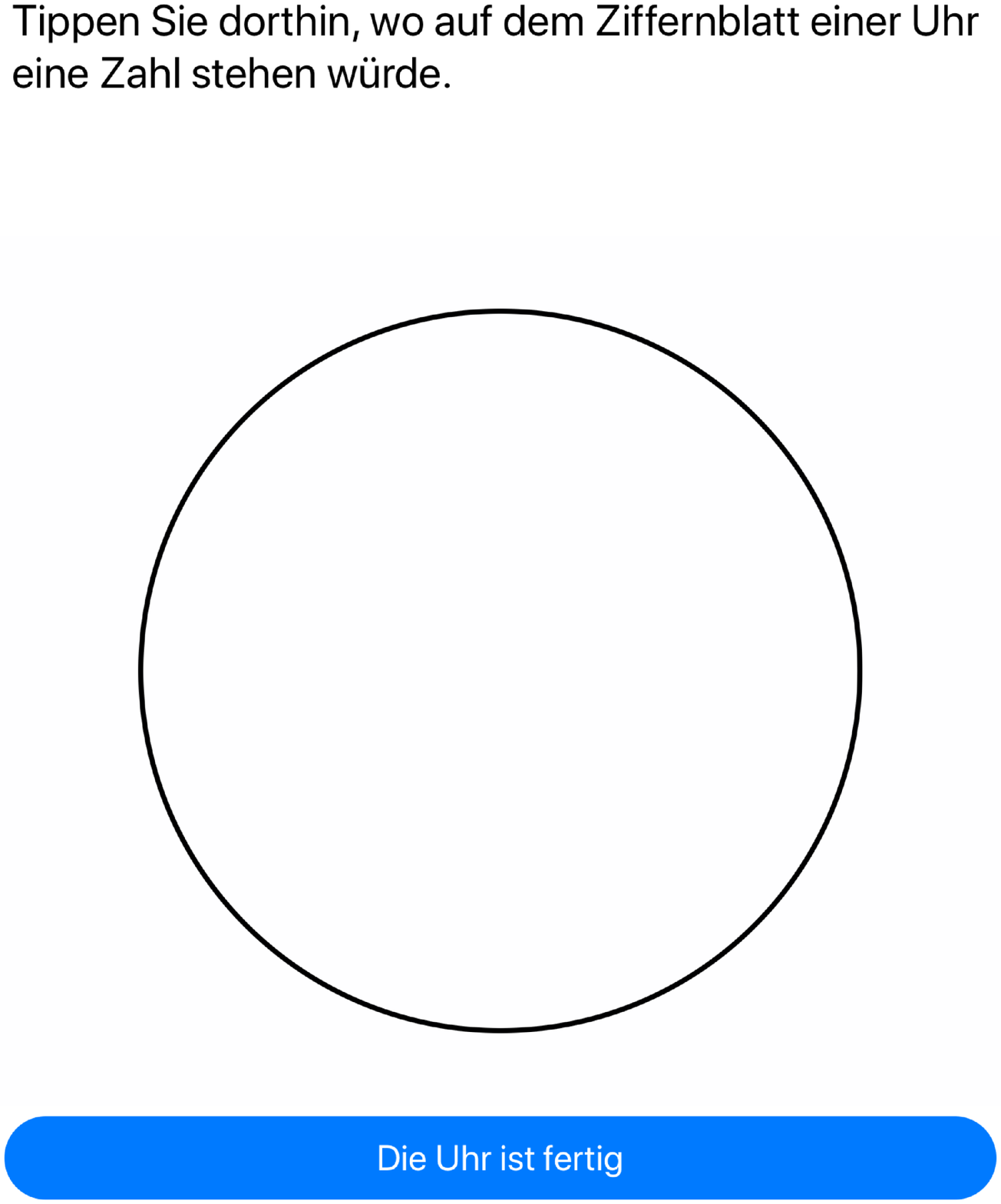}
  \begin{center}
  {\footnotesize(a) Tap on a location in or around the circle.\par}
  \end{center}
\endminipage\hfill
\minipage[t]{0.3\textwidth}
  \includegraphics[width=\linewidth]{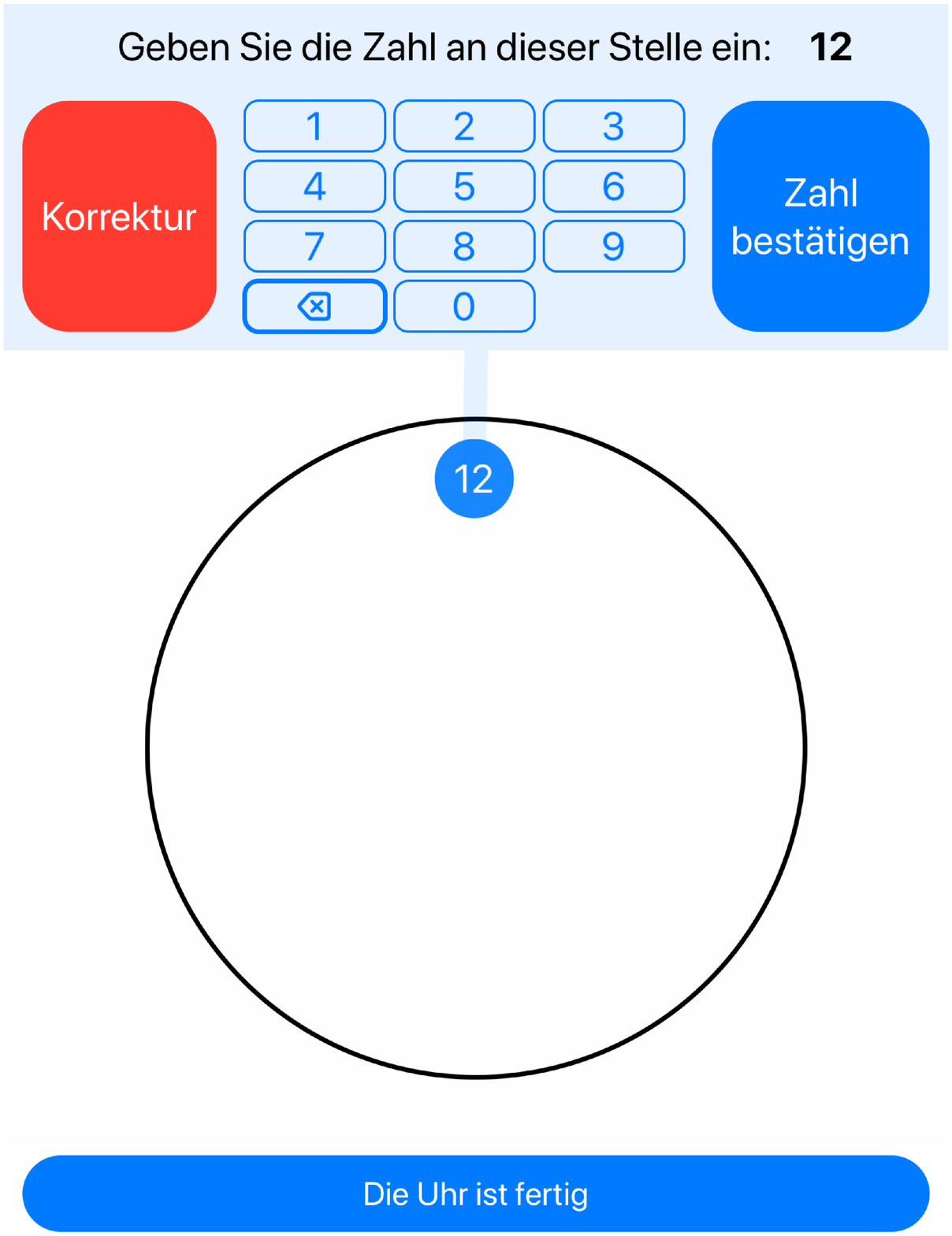}
  \begin{center}
  {\footnotesize(b) Enter a number and confirm it.\par}
  \end{center}
\endminipage\hfill
\minipage[t]{0.3\textwidth}
  \includegraphics[width=\linewidth]{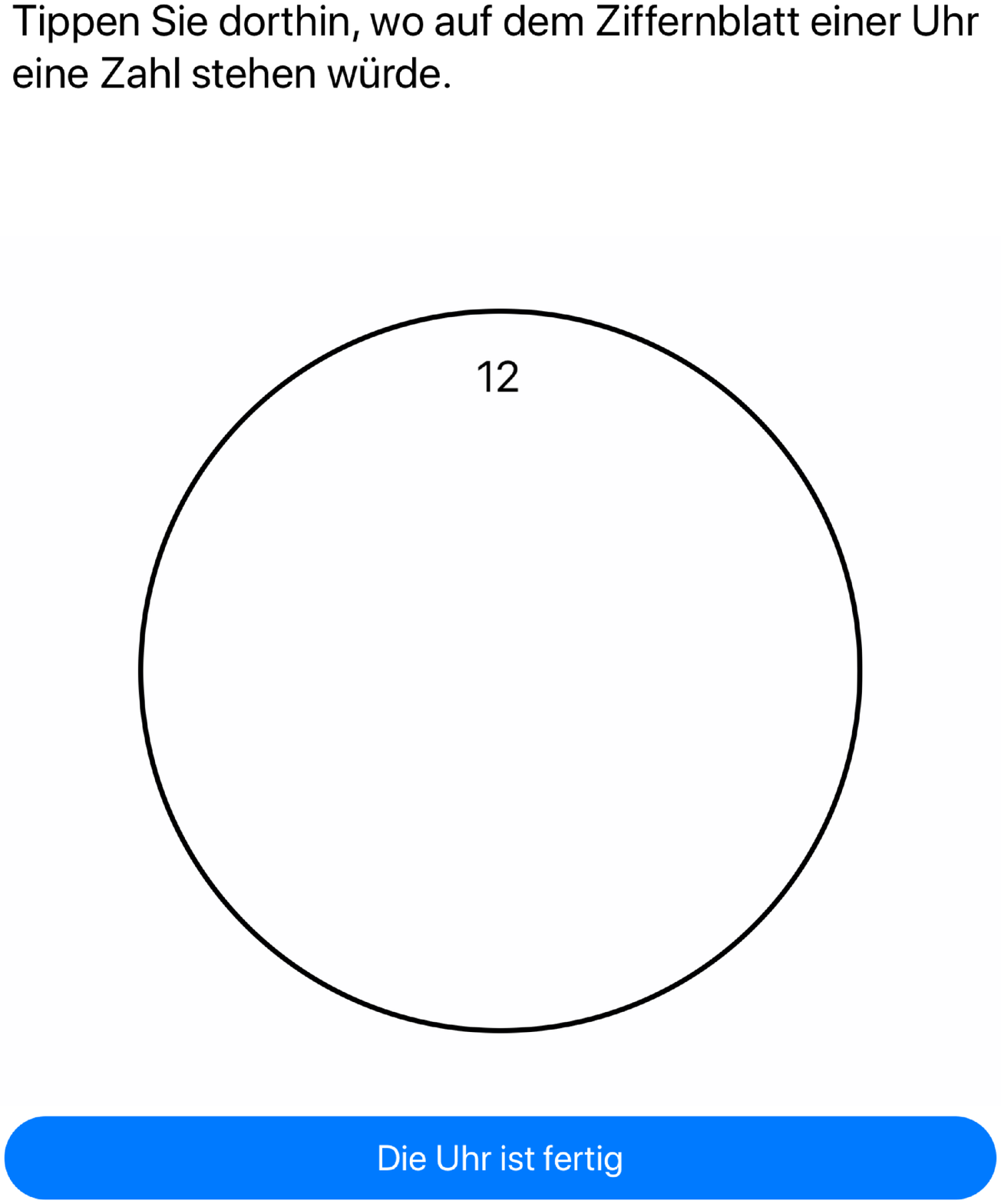}
  \begin{center}
  {\footnotesize(c) The number is added to the clock face.\par}
  \end{center}
\endminipage
\caption{Clock Drawing: steps to add a number to the clock face. See Section \ref{subsubsec:clockDrawing} for a detailed description. Translation of instructions: (a) Tap where a number would be on the face of a clock. (b) Enter the number at this point.}
\label{fig:ClockDrawing_addNumber}
\end{figure}

Clock Drawing is a common subtest in cognitive screening instruments with variations in administration and scoring systems \cite{Ehreke2009}. In DemSelf, all input in Clock Drawing is made by tapping and drawing with a finger. An empty circle is provided as in the Qmci. Numbers and hands can be entered inside the circle and in a quadratic area surrounding it. The subject is first asked to put in all the numbers and then draw the hands into this clock face in a subsequent step. Drawing is done with one finger and creates a straight line between the start and end points.\\
Numbers are added by (1) tapping on a location in the rectangular area (2) entering a number in the appearing number pad (3) confirming the number which closes the number pad (see Figure \ref{fig:ClockDrawing_addNumber}).
A number can be modified by tapping on it – the number pad reappears, and the number can be deleted or changed. When a number is selected this way, it can be relocated by either tapping on a new location or by dragging it to a new location. 

\begin{figure}[!htb]
\minipage[t]{0.32\textwidth}
  \includegraphics[width=\linewidth]{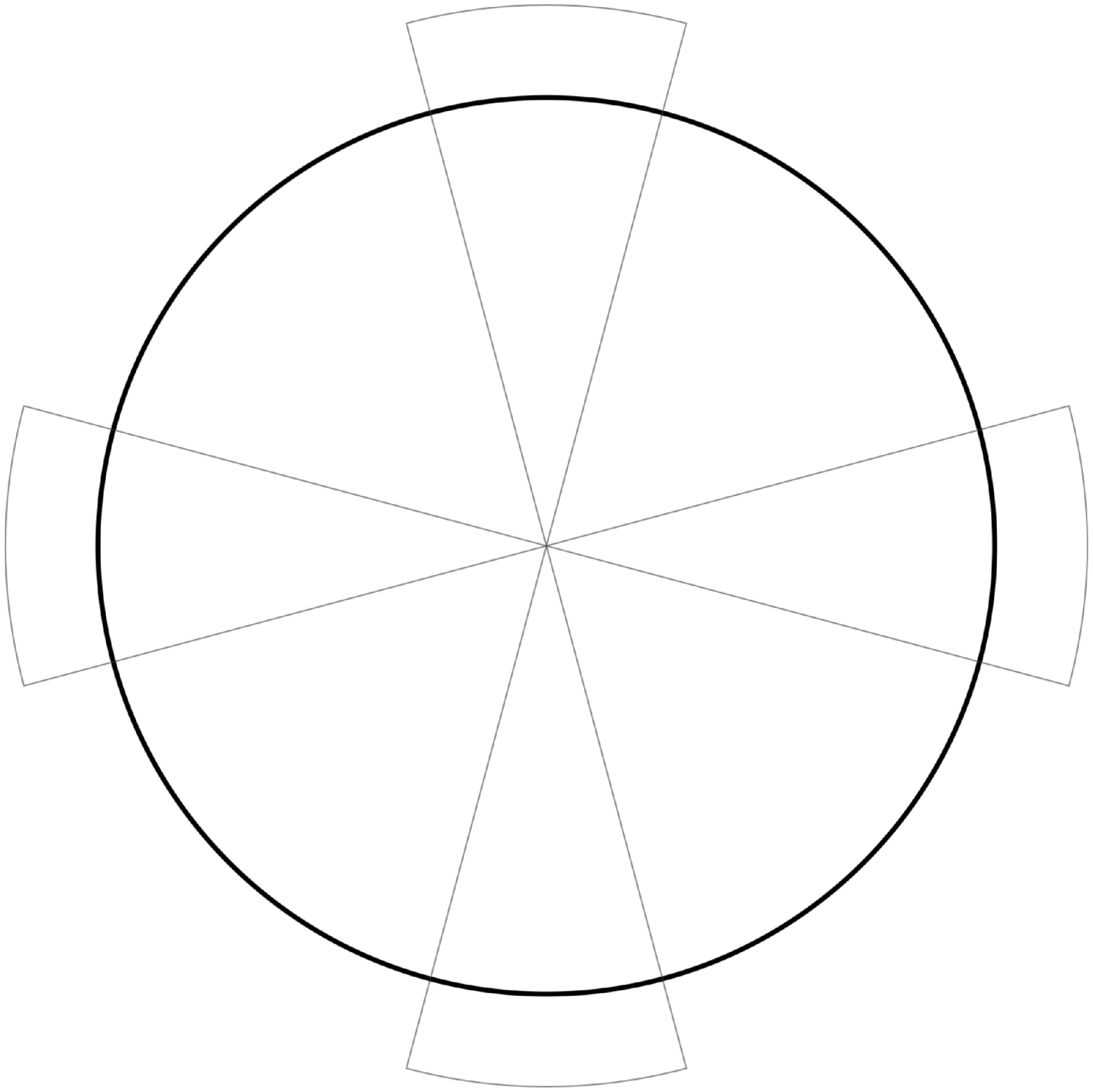}
  \begin{center}
  {\footnotesize(a) Sections for the numbers 12, 3, 6 and 9.\par}
  \end{center}
\endminipage\hfill
\minipage[t]{0.32\textwidth}
  \includegraphics[width=\linewidth]{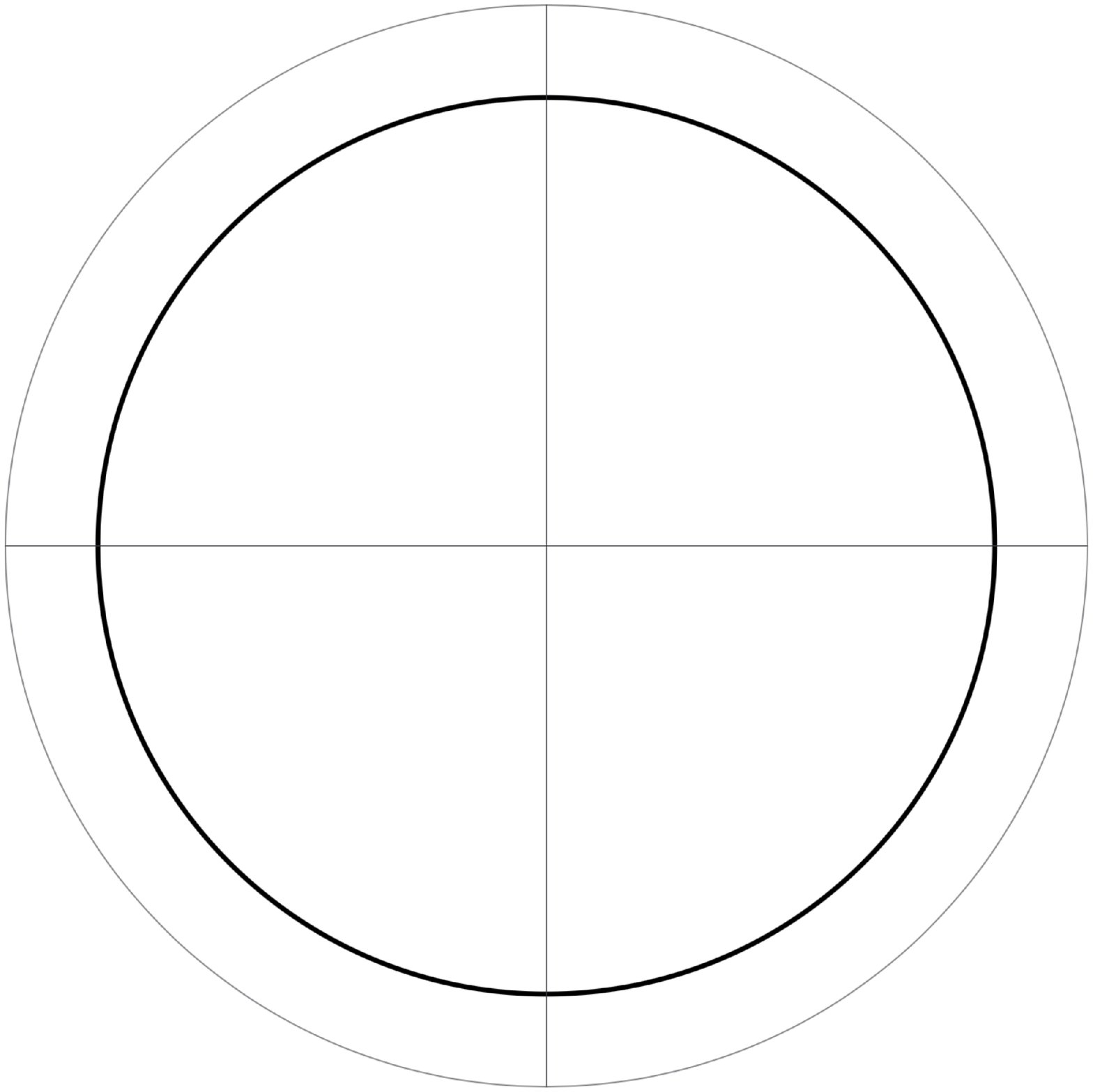}
  \begin{center}
  {\footnotesize(b) Quadrants for the numbers 1,2,4,5,7,8,10 and 11.\par}
  \end{center}
\endminipage\hfill
\minipage[t]{0.32\textwidth}
  \includegraphics[width=\linewidth]{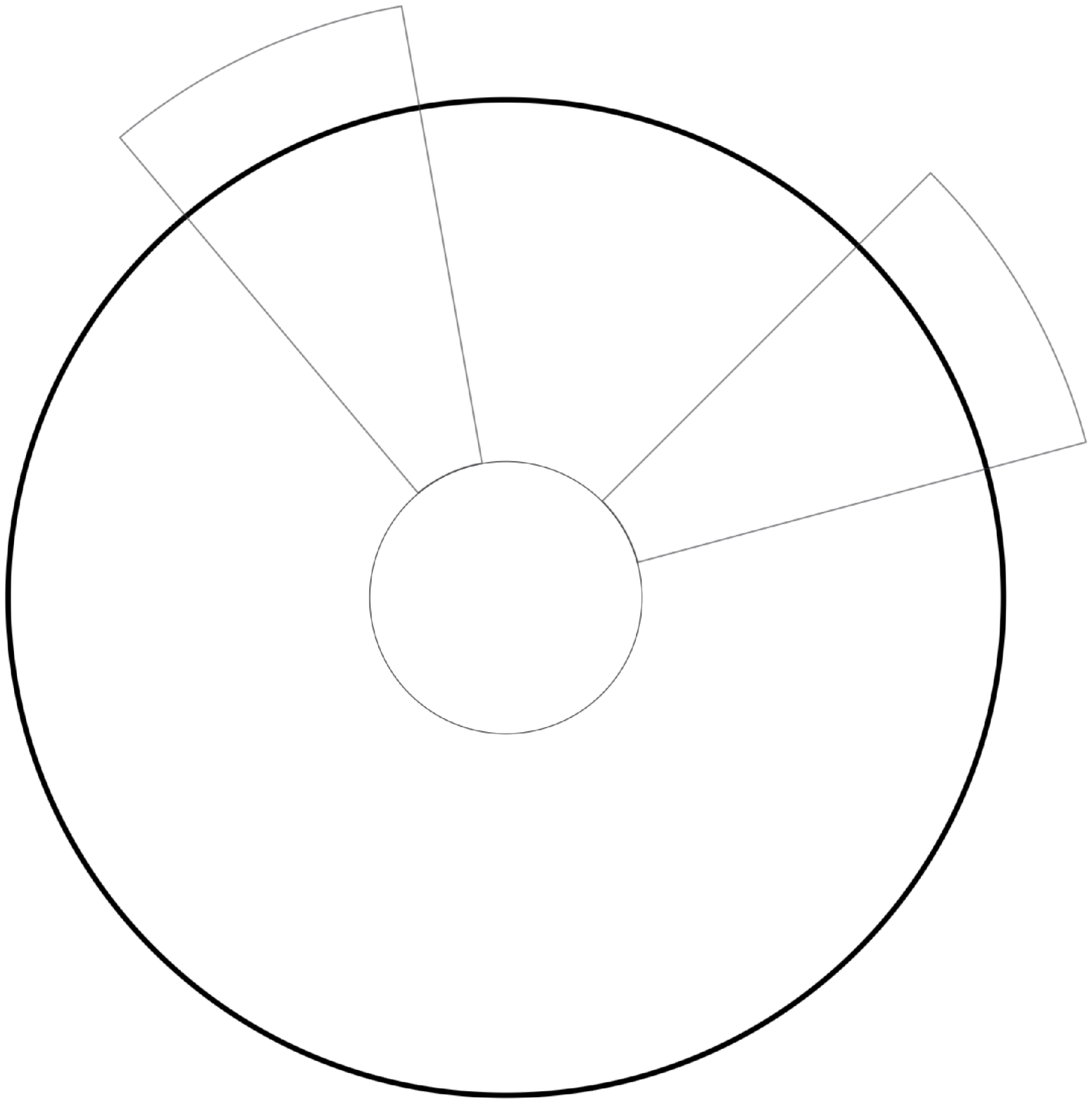}
  \begin{center}
  {\footnotesize(c) Areas for start and end points of the hands.\par}
  \end{center}
\endminipage
\caption{Clock Drawing: target areas for scoring numbers and hands}
\label{fig:ClockDrawing_targetAreas}
\end{figure}

The target areas for numbers and hands are shown in Figure \ref{fig:ClockDrawing_targetAreas}. The numbers 12, 3, 6 and 9 must be located inside a section of 30 degrees and the numbers 1, 2, 4, 5, 7, 8, 10 and 11 must be located inside the corresponding quadrant to be correct. The numbers are also accepted in a limited range outside the circle. Numbers are scored as in the Qmci \cite{Molloy2017}: 1 point for each correct number and minus 1 point for each number duplicated or if greater than 12. Incorrectly placed numbers which are not duplicates or greater than 12 are ignored.

Hands need to be drawn at 10 minutes past 11. Drawing with the finger creates a straight line between start and end point. A hand is considered correct if its end point lies within one of the hand sections shown in Figure \ref{fig:ClockDrawing_targetAreas}. An additional point is given, if both inner end points of the hands lie within the inner circle.

A subject could rotate the device while entering the numbers and hands. The Qmci guide asks the human evaluator to maximize the score by lining up the scoring template at "12 o'clock" \cite{Molloy2017}. In DemSelf, the target areas are rotated and lined up at the location of the number 12 (if 12 is missing, 3, 6, or 9 are used respectively). The maximum of the unaligned or aligned score is selected.

\subsubsection{Subtest Delayed Recall}
In Delayed Recall, test subjects are asked to repeat the words that have been presented in the subtest Word Registration. Again, there are 16 answer choices with 11 randomly selected distractors.

\subsubsection{Subtest Logical Memory}
\label{subsubsec:logicalMemory}
Logical Memory tests the recall of a short story. Instead of free verbal recall, repeating the story is divided into 9 steps. For each step, there are 6 answer choices with the correct story component and 5 distractors. The distractors were chosen to (1) have the same syntactic structure, (2) be semantically related, and (3) be used with a similar frequency in German as the original item. A new story component is added to the answer by tapping on the according button. The current answer is displayed on the screen.

\begin{figure}[ht]
\minipage[t]{0.3\textwidth}
  \includegraphics[width=\linewidth]{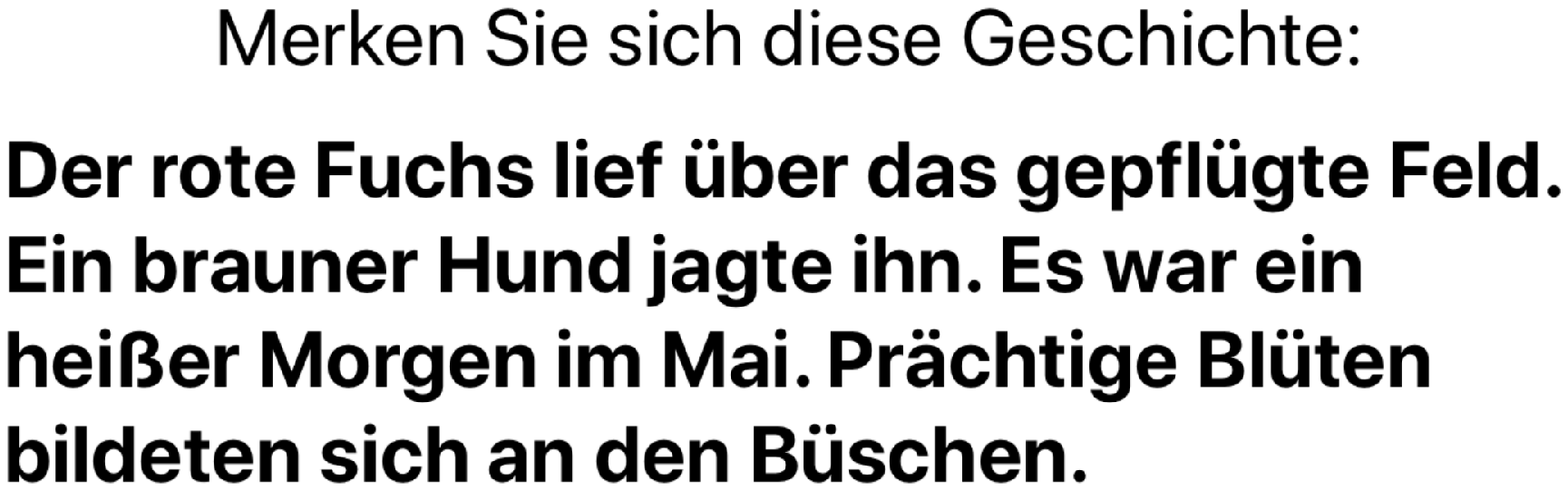}
  \begin{center}{\footnotesize(a) The story is displayed and read aloud.\par}\end{center}
\endminipage\hfill
\minipage[t]{0.3\textwidth}
  \includegraphics[width=\linewidth]{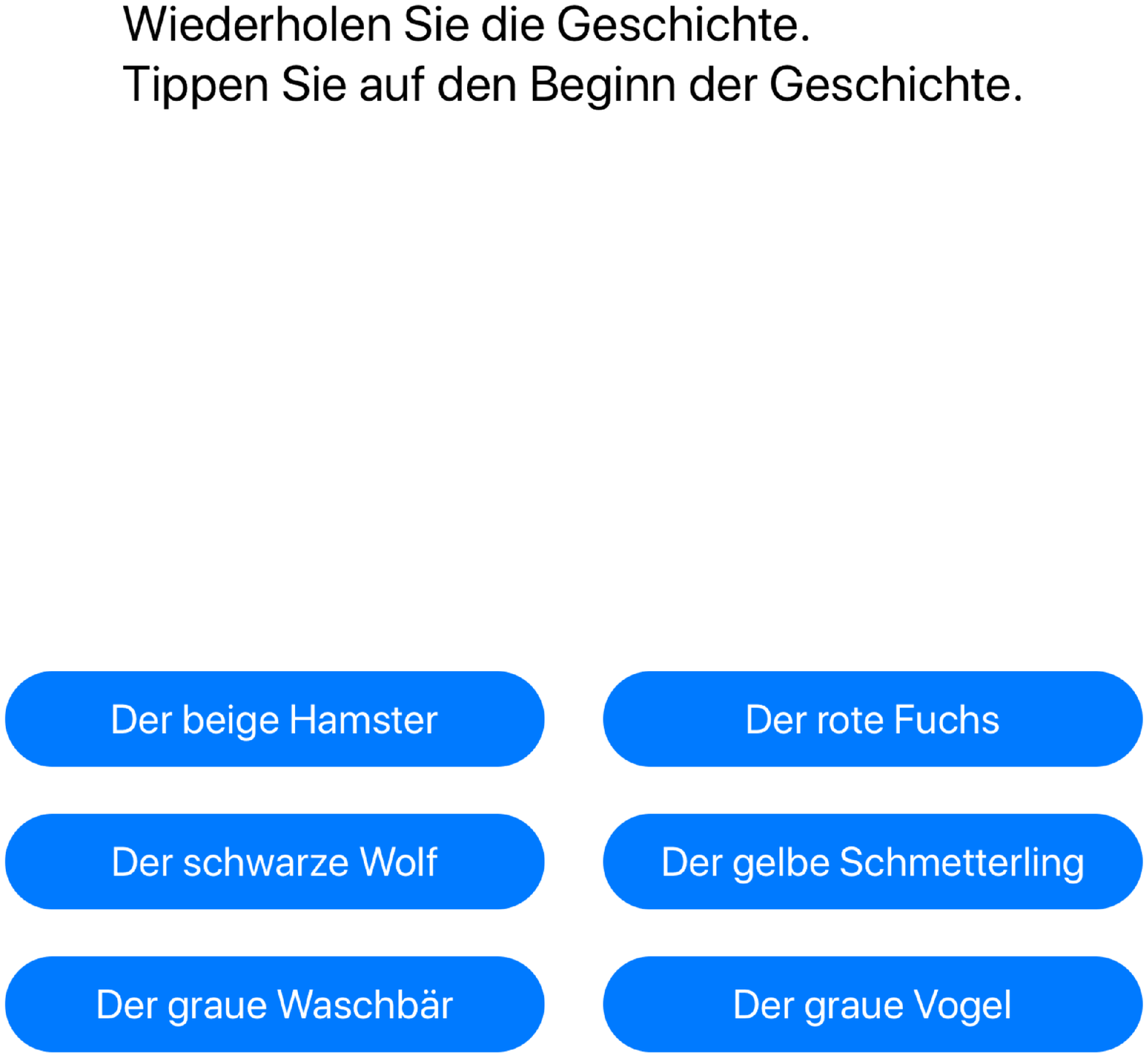}
  \begin{center}{\footnotesize(b) Subject taps the beginning of the story.\par}\end{center}
\endminipage\hfill
\minipage[t]{0.3\textwidth}
  \includegraphics[width=\linewidth]{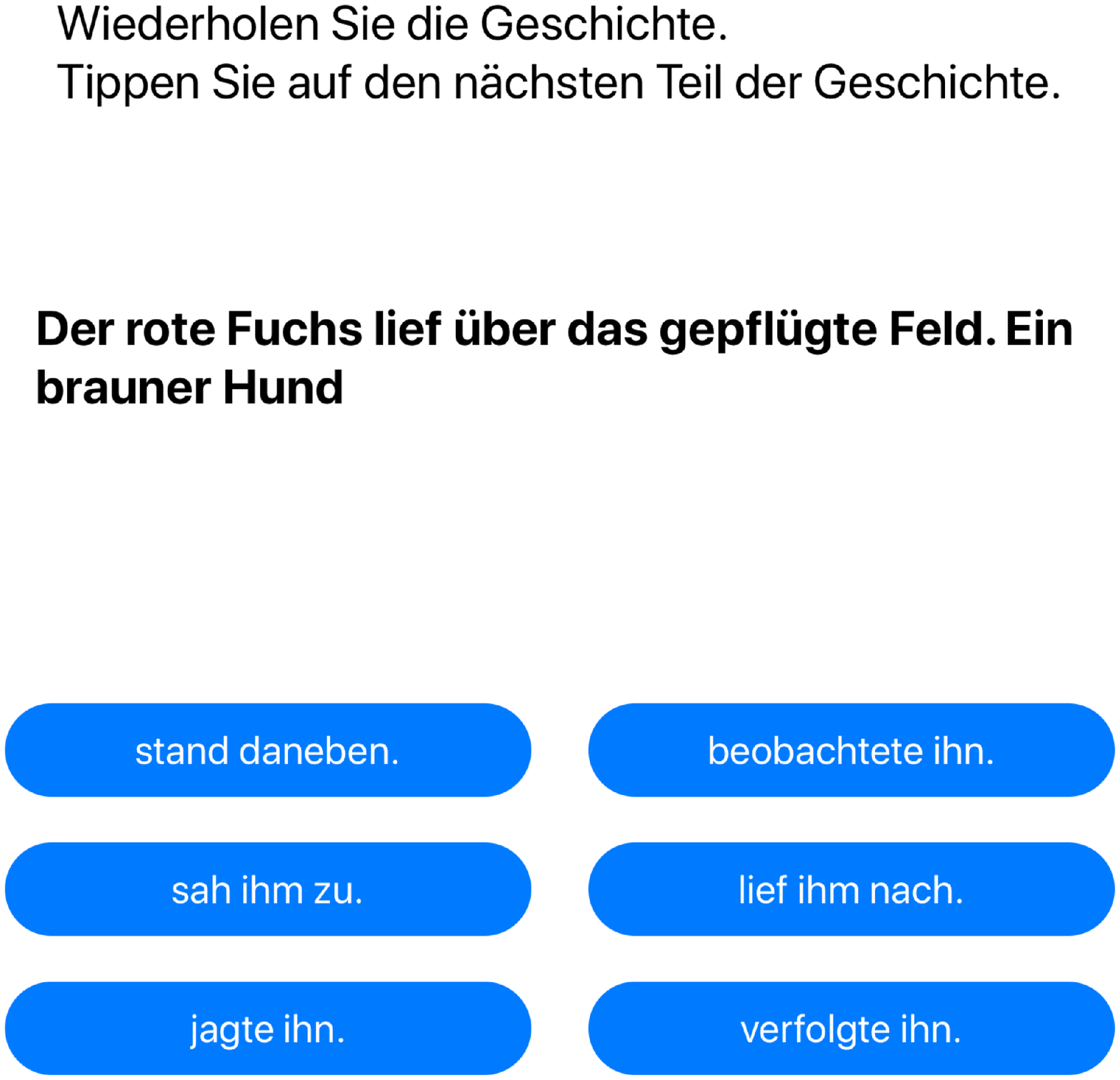}
  \begin{center}{\footnotesize(c) The story is completed step by step.\par}\end{center}
\endminipage
\caption{Logical Memory: Item presentation and repeating a story in several steps. See Section \ref{subsubsec:logicalMemory} for a detailed description. Translation of instructions: (a) Remember this story. (b) Repeat the story. Tap the beginning of the story. (c) Repeat the story. Tap the next component of the story.}
\label{fig:LogicalMemory_input}
\end{figure}

\subsubsection{Test Result}
It is reported to the subject whether there is evidence of normal cognition or risk for cognitive impairment. Subjects are once again encouraged to speak with a physician if they have any concerns. Test results can be sent to the physician who initiated the testing or saved on the tablet.

A more detailed test result is intended for medical professionals. The test score and the cut-off scores are displayed on a bar plot. We used the original cut-off scores from Qmci, but scaled down because the Verbal Fluency subtest was missing. Additional information such as the test date, start time and duration, as well as the subject's age and education are displayed. The subject's answers about the test environment and current physiological and emotional status are also reported. Further details can be displayed for each subtest, including the subject's responses, an explanation for the subtest's scoring system, and additional information such as the time taken to complete the subtest.

\section{Results}
Below we present some of the feedback on the DemSelf instrument that was mentioned by two or more participants or was considered particularly important.

\subsection{Usability}
Participants identified usability issues in all components of the DemSelf app. In this section, we present a selection of important usability issues noted by two or more participants.\\
Participants generally supported the idea to test a subject’s understanding that the test result alone does not provide a diagnosis and can be inaccurate. However, hiding the button to proceed to the next screen until the user has answered correctly without providing any feedback was considered bad practice (see Figure \ref{fig:ConsentTestEnvironmentScreens}).
Leaving the application to make adjustments in the device settings or to get more information about cognitive impairments in the browser may cause problems, as subjects may not be able to find their way back. In addition, reading about dementia and commonly used cognitive tests could lead to priming effects that influence behavior on the test.
The buttons for changing volume on the iPad might be difficult to find for subjects unfamiliar with tablets. Participants suggested displaying arrows on the screen pointing to the volume buttons.
Participants also missed an indicator of how many screens are displayed before the actual test begins.\\
The confirmation of answers in the subtests was not consistent. Responses are sometimes logged directly when tapping a button and can no longer be changed, such as in the Orientation and Logical Memory subtests.
This can be particularly frustrating for people with vision or motor impairments who accidentally touch the wrong button. Such operating errors also lead to bias within the test result.
In the Word Registration and Delayed Recall subtests, the screen changes automatically when five words have been logged in, so that the fifth word cannot be changed. Previous answers can be deselected by tapping on them. This behavior was irritating and inconsistent. Participants suggested adding a global confirmation button to all subtests that logs in the selected answers and switches to the next screen.
In the Orientation subtest, the next question appears after a time limit of 10 seconds, which was considered a serious usability problem. Subjects may feel they are loosing control. As a result, some participants also falsely assumed a time limit in the following subtests.
Scores may still depend on reaction time, but the end of a time limit should not result in an automatic screen change.
When items are presented both in text form and verbally, new text passages should be displayed step-by-step and synchronously with the spoken word, as the speed of reading and following the verbal presentation may differ.\\
In Clock Drawing, almost all participants initially overlooked the buttons to confirm a number or hand and add it to the clock. Instead, participants tapped a new location in the clock circle or began drawing a new hand. The confirmation process intended by the system was unexpected, since no input had to be explicitly confirmed in the previous subtests.
Numbers can be relocated on the clock by drag-and-drop, which was perceived as intuitive for younger people. However, dragging a number across the screen requires fine motor skills that many elderly people do not have. The mechanism also was not explicitly mentioned, which could influence the test result.
A subject's arm and hand will partially cover the clock while using the number pad (see Figure \ref{fig:ClockDrawing_addNumber}). A position below the clock circle would prevent this and correspond to the common conventions for mobile devices.

\subsection{Adaption}
DemSelf is a different test than Qmci and needs to be re-validated. This section presents important changes that were introduced by the adaption as self-administered touch-based instrument.
Visual recognition and selection between several answer options instead of a free verbal recall is one of the main differences between the Qmci and DemSelf, with implications on the involved cognitive processes and the subtest difficulty. 
In all subtests besides Clock Drawing, users provide answers by tapping on labeled buttons and no longer have to verbally recall the items. This changes the nature of the task.
Overall, there was no clear consensus on how many or what type of distractors are appropriate.\\
Reading skills, visual memory abilities, or visual impairments are new confounding variables in DemSelf. Items are presented in text form for a certain period, and a fast reader can scan the words several times. 
The same applies to labeled answer buttons when there is a time limit, such as in the Orientation subtest. Different questions in Orientation will likely require different time limits as different amounts of text are displayed.
We reported in the previous section that people with non-cognitive impairments, such as tremor or visual impairments, could unintentionally give wrong answers by tapping on wrong buttons. Participants therefore raised doubts about whether Orientation is valid and really tests whether subjects know the answer.\\
For the Clock Drawing subtest, participants reported that important aspects from the paper-based version such as spatial orientation and understanding the clock are also tested in the adapted version. Many typical errors also seemed possible in the adapted version. Experience with touchpads and mobile devices could be an important factor for this subtest, as some interaction options are not explicitly mentioned, such as dragging a number to change the position.
In the paper-based version of Clock Drawing, it is common to draw auxiliary lines or a dot in the middle of the clock before filling in the numbers. This is not possible in DemSelf, nor is drawing curved hands.\\
Scoring in Logical Memory is based on the number of target words recalled, which makes sense for free verbal recall. However, in the adapted version, a correct story component with two target words such as "Der rote Fuchs" [The red fox] may actually be easier to remember than a story part like "im Mai." [in May] with only one target word, and thus should not score higher.

\subsection{Acceptance}
Participants were generally cautious about self-assessment in practical application. Participants assumed that many older people would not be able to self-administer the test due to a lack of technical experience. However, as mobile devices are increasingly used by older people, this problem will decrease in the future.
For now, an assisting person should be present at least to start the app and to adjust volume and brightness.
Participants would generally prefer if the test is conducted in a medical office rather than at a patient's home so that someone is present for support, and the result can be discussed directly.
Healthcare professionals may also hesitate to use DemSelf because they lack information from observation and interaction with the patient. 
It was therefore positively noted that the state of the subject is reported in the test result as fatigue and pain can affect the test result, which means that the subject should be retested.
Reporting the test result directly to the subject in the app was viewed very negatively. It would not be ethical to leave individuals alone with an unexpected test result about MCI or dementia that may also be incorrect. 
DemSelf could also be used in a traditional context with a human administrator to automatically score a test and save the results. Given the current trend towards digitizing patient data, automatic scoring and digital test results are beneficial for clinical practice.
Because DemSelf is a different test than Qmci, participants agreed that DemSelf needs validation to be used in practice. Validation must also clarify which requirements a test person must fulfil in order to use DemSelf. Subjects must at least be able to read and must not have a severe visual or motor impairment. For people who can perform the test, DemSelf is a promising direction for application and research.

\section{Discussion}
Developing a self-assessed cognitive screening instrument is challenging. Psychometric properties must be ensured, taking into account a variety of technical and human factors \cite{Bauer2012}. This study provides some insights for the development, validation and practical application of self-administered cognitive screening instruments. DemSelf is still a rather traditional instrument as it is directly based on a validated paper-based instrument. Some potential benefits of computerized tests, such as adapting to individual performance or the use of machine-learning algorithms to interpret test performance \cite{Yim2020,Sternin2019,Cha2019}, are not part of the instrument.
A high level of usability is a prerequisite for successful practical use. Test subjects should immediately feel confident in using the device, as the test takes only a few minutes and is not repeated. However, it is in the nature of a standardized cognitive test that certain usability heuristics are hard to fulfill. Items can, for example, be presented only once because otherwise the test results would no longer be comparable. This rigidity can also be a problem for computerized tests.
The confirmation of inputs should be handled consistently throughout the subtests. Subjects must be able to change an answer, just as in a human-administered test, where a subject can revoke a statement or cross out a number or hand. Errors that are not due to cognitive impairment, such as manually missing the right button should be avoided.
An open question is still to what extent the measurement of time limits can be used as in the Qmci. Automatic change of screens after 10 seconds during orientation questions was one of the most criticized usability issues. Even if the screens do not change automatically, differing reading abilities and visual or motor impairments add a lot of variance to the time it takes to answer a question that is not related dementia or MCI.\\
One of the main differences between the original Qmci and DemSelf was recall vs. recognition.
A comparison of free recall and either yes/no recognition or three-alternative forced-choice recognition in cognitively-impaired subjects found that yes/no recognition was the best predictor of MCI and early AD \cite{Bennett2006}. Yes/no recognition could therefore be used in an updated version of DemSelf.
A digital pen was suggested for the Clock Drawing subtest to make it more similar to the original paper-based version. \cite{SouillardMandar2016} successfully used such a digital pen as well as a machine learning approach for evaluation of the Clock Drawing test. In the DemSelf implementation, users only need to tap a certain location to enter numbers and can only draw straight lines. The reduced vulnerability to motor impairments in the DemSelf implementation could be a strength of the instrument.
User testing is required to determine if DemSelf is appropriate for certain subpopulations – for example, for those who are uncomfortable using a tablet or have certain limitations such as aphasia, poor reading skills, or hemiparesis \cite{Bauer2012}. 
Some factors can affect a self-assessed computerized screening instrument differently than an examiner-administered paper-based instrument. 
For example, technology use and commitment towards technology has been shown to affect test scores when the same test is administered on paper or on a tablet \cite{Steinert2018}. 
Technical experience, reading skills, and motor impairments were mentioned as possible confounding factors for DemSelf. \\
The expert interviews revealed possibilities for improving the prototype. By completing the test twice, participants could explore different ways of interaction and discover usability problems that would otherwise have gone unnoticed. 
The involvement of stakeholders from dementia and MCI research allowed us to evaluate the acceptability of the instrument for practical use and to identify usability issues that are specific to elderly and cognitively-impaired users.
In a future study, screen capture or video recordings can help to better analyze the interaction with complex tasks like the clock drawing test. 
One limitation of this study is that no users from the target group were involved. Until then, the presented comments on usability remain partly speculation. Similar to \cite{Zeng2018} the next step in the evaluation of DemSelf could be focus group interviews with healthy and cognitively-impaired elderly users. A study comparing the test performance in Qmci and DemSelf would shed more light on the discussed differences between the instruments. In the future, we hope to allocate more funding in order to further our research on the usability and validity of computerized cognitive screening instruments.

\section{Conclusion}
Early detection of MCI and dementia is vital as many therapeutic and preventive approaches are particularly effective at an early stage. Cognitive screening instruments are primarily paper-based and administered by healthcare professionals. Computerized cognitive screening instruments offer a number of potential advantages over paper-based instruments, such as increased standardization of scoring, reduced costs, remote testing and a more precise measurement of time- and location-sensitive tasks \cite{Bauer2012}.\\
We presented a touch-based cognitive screening instrument, called DemSelf. It was developed by adapting an examiner-administered paper-based instrument, the Quick Mild Cognitive Impairment (Qmci) screen.
Usability is a key criterion for self-assessment by elderly users with potential cognitive and other impairments. We conducted interviews with experts in the domain of usability and human-machine interaction and/or in the domain of dementia and neuropsychological assessment to evaluate DemSelf.
The expert interviews revealed possibilities for improving the prototype.
Developers should consider barriers for elderly and inexperienced users, such as not being able to reverse an answer or leaving the application without finding their way back. Time limits cannot be taken directly from a paper-based test version and should not lead to automatic screen changes.
Visual recognition instead of a free verbal recall is one of the main differences between the Qmci and DemSelf. Reading skills, technical experience, and motor impairments seem to be important confounding variables. Further research is needed to determine validity and reliability of computerized versions of conventional paper-based tests.
Healthcare professionals may hesitate to use DemSelf because they lack information from observation and interaction with the patient – that could also be evaluated electronically in the long-term.
Participants would generally prefer if the test is conducted in a medical office rather than at a patient's home so that someone is present for support and the result can be discussed directly.
In view of the current trend towards digitization of patient data, automatic scoring and digital test results are beneficial for clinical practice. Computerized instruments like DemSelf also have the potential to reduce costs and provide early support for people with MCI and dementia.

\bibliographystyle{splncs04}
\bibliography{references.bib} 

\end{document}